\documentclass[journal]{IEEEtran}

\newtheorem{lemma}{Lemma}

\newtheorem{remark}[lemma]{Remark}
\newtheorem{proposition}[lemma]{Proposition}
\newtheorem{definition}[lemma]{Definition}
\newtheorem{theorem}[lemma]{Theorem}
\newtheorem{corollary}[lemma]{Corollary}

\providecommand{\qed}{\hfill\rule{1.5ex}{1.5ex}}

\DeclareMathAlphabet{\myvec}{OML}{cmm}{b}{it}

\begin{document}
%
\title{Primal-dual distance bounds of linear codes
with application to cryptography}
%
%
\author{Ryutaroh~Matsumoto,~\IEEEmembership{Member,~IEEE,}
        Kaoru~Kurosawa,~\IEEEmembership{Member,~IEEE,}
        Toshiya~Itoh,~\IEEEmembership{Nonmember,}
        Toshimitsu Konno,~\IEEEmembership{Nonmember,}
        Tomohiko Uyematsu,~\IEEEmembership{Member,~IEEE}
\thanks{To appear in IEEE Transactions on Information Theory, Sept.\ 2006.
        This work was supported by the Okawa Foundation
for Information and Telecommunications, Japan.}
\thanks{R.~Matsumoto and T.~Uyematsu are with
the Department of Communications and Integrated Systems,
Tokyo Institute of Technology,
2--12--1 O-okayama, Meguro-ku, Tokyo 152--8552, Japan.
{\tt \{ryutaroh,uematsu\}@it.ss.titech.ac.jp}}%
\thanks{K.~Kurosawa is with
the Department of Computer and Information Sciences,
Ibaraki University,
4-12-1 Nakanarusawa, Hitachi, Ibaraki, 316-8511, Japan.
{\tt kurosawa@mx.ibaraki.ac.jp}}%
\thanks{T.~Itoh is with
the Global Scientific Information
and Computing Center (GSIC),
Tokyo Institute of Technology,
2--12--1 O-okayama, Meguro-ku, Tokyo 152--8552, Japan.
{\tt titoh@dac.gsic.titech.ac.jp}}%
\thanks{T.~Konno is with
the Department of Computer Science,
Tokyo Institute of Technology,
2--12--1 O-okayama, Meguro-ku, Tokyo 152--8552, Japan.
{\tt tkonno@saturn.sannet.ne.jp}}}%
\maketitle

\begin{abstract}
Let $N(d,d^\perp)$ denote the minimum 
length $n$ of a linear code $C$ with $d$ and $d^{\bot}$,
where $d$ is the minimum Hamming distance of $C$
and
$d^{\bot}$ is the minimum Hamming distance of $C^{\bot}$.
In this paper,
we show a lower bound and an upper bound on $N(d,d^\perp)$.
Further,
for small values of $d$ and $d^\perp$, we determine
$N(d,d^\perp)$ and give a 
generator matrix of the optimum linear code.
This problem is directly related to the design method of cryptographic
Boolean functions suggested by Kurosawa et al.
\end{abstract}


%
\IEEEpeerreviewmaketitle

\section{Introduction}



One of the fundamental problems in coding theory
is to find the minimum length of linear codes
for the given minimum Hamming distance $d$
and the given number of codewords $K$,
where the length of a linear code means
the length of the codewords.

In this paper,
we study a variant of this problem:
find the minimum length of linear codes $C$
which achieves 
the given minimum Hamming distance $d$ 
and
the given minimum Hamming distance $d^{\bot}$ of $C^{\bot}$,
where $C^{\bot}$ denotes the dual code of $C$.
Note that the number of codewords $K$
is replaced by 
the minimum Hamming distance $d^{\bot}$ of $C^{\bot}$
in our new problem.
This problem is interesting not only theoretically
but also practically:
it is directly related to the design of cryptographic
Boolean functions as follows.

Block ciphers must be secure against various attacks,
in particular against differential attacks \cite{biham90}
and linear attacks \cite{matsui93}.
The security of block ciphers is often studied by viewing
their S-boxes (or $F$ functions) as a set of Boolean functions.
We say that
a Boolean function $f(\myvec{x})$ satisfies
(propagation criteria) $PC(\ell)$ \cite{Pre91,Pre}
if $f(\myvec{x})+f(\myvec{x}+\Delta)$
is uniformly distributed for any $\Delta$
with
$1 \leq wt(\Delta) \leq l$,
where  $wt(\Delta)$ denotes the Hamming weight of $\Delta$.

It is clear that
$PC(\ell)$ is directly related to 
the security against differential attacks
because
 $\Delta$ is the input difference
and $f(\myvec{x})+f(\myvec{x}+\Delta)$
is the output difference of $f$.
Also,
$f(\myvec{x})$ is a bent function \cite[Chapter 14]{macwilliams77} if and only if 
$f(\myvec{x})$ satisfies $PC(n)$ \cite{Pre},
where
a bent function has the largest distance from the set 
of affine (linear) functions.
Hence $PC(n)$  is directly related to the security
against linear attacks.
The famous strict avalanche criterion (SAC),
which was introduced
as a criterion of the security of  S-boxes \cite{SAC},
 is equivalent to $PC(1)$.

More generally,
we say that
$f(\myvec{x})$ satisfies
(extended propagation criteria)  $EPC(\ell)$ of order $k$ \cite{Pre91,Pre}
if
$f(\myvec{x})$ satisfies $PC(\ell)$ even if 
any $k$ bits of $\myvec{x}=(x_1, \cdots, x_n)$
are fixed to any constant bits.
(We remark that many authors refer to EPC as just PC, 
including \cite{kurosawa97}.)
For example,
SAC$(k)$,
which is a generalized version of SAC,
 is equivalent to $EPC(1)$ of order $k$.
As shown above,
$EPC(\ell)$ of order $k$ is a more generalized
security notion of cryptographic Boolean functions.

Kurosawa et al. \cite{kurosawa97}
gave the first construction method
of $EPC(\ell)$ of order $k$
based on the Maiorana-McFarland construction
(see \cite{dillon74}).
They showed that
there exists an $EPC(\ell)$ of order $k$ function
$f(x_1, \cdots, x_{n})$ 
if there exists a linear code $C$
such that $d=k+1$, $d^{\bot}=\ell+1$
and
the length of $C$ is $n/2$,
where $d$ is the minimum Hamming distance of $C$
and
$d^{\bot}$ is the minimum Hamming distance of $C^{\bot}$.
Carlet generalized this construction
to nonlinear codes \cite{carlet99}.

We now ask, given $k$ and $\ell$,
what is the minimum $n$
for which $EPC(\ell)$ of order $k$ functions
$f(x_1, \cdots, x_n)$ exist ?
In the design method of
Kurosawa et al. \cite{kurosawa97},
this is equivalent to saying that,
given $d$ and $d^{\bot}$,
find the minimum length $n$ of a linear code $C$ with $d$ and $d^{\bot}$.
Note that
this problem is exactly the same as the one 
mentioned at the beginning of the introduction.

More formally,
let $N(d,d^\perp)$ denote the minimum 
length $n$ of a linear code $C$ with $d$ and $d^{\bot}$,
where $d$ is the minimum Hamming distance of $C$
and
$d^{\bot}$ is the minimum Hamming distance of $C^{\bot}$.
We then want to find $N(d,d^\perp)$ 
for given $d$ and $d^{\bot}$.
In this paper,
we show lower bounds and upper bounds on $N(d,d^\perp)$.
Further,
for small values of $d$ and $d^\perp$, we determine
$N(d,d^\perp)$ exactly and give a generator matrix of 
the optimum linear code.

This paper is organized as follows:
In Section 2, we introduce relevant concepts and notations.
In Section 3, we propose upper bounds on $N(d,d^\perp)$.
In Section 4, we propose lower bounds on $N(d,d^\perp)$,
show true values of $N(d,d^\perp)$,
and compare the proposed bounds with the true values.
In Section 5, concluding remarks are given.


\section{Preliminaries}


\subsection{Notation}


We use $f$ to denote a Boolean function 
$\{0,1\}^n \rightarrow \{0,1\}$,
and $\phi$ to denote a function 
$\{0,1\}^n \rightarrow \{0,1\}^m$, where $m\leq n$.
We use $\myvec{x}$ to denote $(x_1, \cdots,x_n)$,
where $x_i$ is a binary variable.

Let $\cdot$ denote the inner product of two binary vectors
over $GF(2)$.
For a set $A$,
$|A|$ denotes the cardinality of $A$.

Let a linear $[n,m,d]$ code denote a binary linear code $C$
of length $n$, dimension $m$
and 
the minimum Hamming distance at least $d$.
The dual code $C^{\bot}$ of a linear code $C$ is defined as
$C^{\bot} = \{ u \mid u \cdot v =0 \mbox{ for all } v\in C\}\enspace$.
The dual distance $d^\bot$ of $C$
is defined as the minimum Hamming distance of $C^\bot$.


\subsection{Resilient Functions}


\begin{definition}
We say that
$\phi:\{0,1\}^n \rightarrow \{0,1\}^m$ is an $(n,m,k)$-resilient function
if $\phi(x_1, \cdots, x_n)$ is uniformly distributed 
even if any $k$ variables 
$x_{i_1}, \cdots, x_{i_k}$ are fixed into constants.
That is,
\[   \Pr [\phi(x_1,\ldots, x_n)=(y_1,\ldots,y_m) \mid
           x_{i_1}x_{i_2} \cdots x_{i_k}=\alpha ] = 2^{-m}  \]
for any $k$ positions $i_1  < \cdots < i_k$,
for any $k$-bit string $\alpha\in  \{0,1\}^k $
and for any fixed
$(y_1, \cdots, y_m) \in \{0,1\}^m$, where the values 
$x_j$ ($j \not\in \{i_1, \dots , i_k\}$) are chosen independently at random. 

\end{definition}


\subsection{$EPC(\ell)$ of order $k$}


Define the derivative of $f:\{0,1\}^n \rightarrow \{0,1\}$ 
by
\[ D_{\Delta} f=f(\myvec{x}) +f(\myvec{x}+\Delta) \]
for $\Delta \in \{0,1\}^n$.

\begin{definition}
\cite{Pre91,Pre}
We say that a Boolean function
$f:\{0,1\}^n \rightarrow \{0,1\}$ 
satisfies $EPC(\ell)$ of order $k$ if $D_{\Delta} f$ is $k$-resilient 
for any $\Delta \in \{0,1\}^n$ 
with $1\leq wt(\Delta) \leq l$.
(We also say that
$f$ is an $EPC(\ell)$ of order $k$ function.)
\end{definition}




Kurosawa et al.
gave a general method to design
$EPC(\ell)$ of order $k$ functions by using
a linear code \cite{kurosawa97}.

\begin{proposition}\label{boolean}
\label{pro:pccode}
Suppose that there exists
 a linear $[n,m,k+1]$ code $C$ 
with the dual distance at least $\ell+1$.
Then
there exists 
an $EPC(\ell)$ of order $k$ function 
$f:\{0,1\}^{2n} \rightarrow \{0,1\}$.
\end{proposition}

\begin{remark}
The construction of \cite{kurosawa97} is essentially
quadratic in nature with a non-quadratic `offset' part.
After \cite{kurosawa97}, Carlet \cite{carlet99} showed a construction which uses nonlinear Kerdock and Preparata codes as an improvement.
It gives
non-quadratic Boolean functions not just in their offset part.
\end{remark}

Define $N(d, d^{\perp})$ as the minimum $n$
such that
there exists 
 a linear $[n,m,d]$ code $C$ 
with the dual distance at least $d^{\perp}$.
Then
$N(k+1, \ell+1)$ is the minimum $n$
such that
there exists 
a $EPC(\ell)$ of order $k$ function 
$f:\{0,1\}^{2n} \rightarrow \{0,1\}$
in the design method of Kurosawa et al.
We will consider the upper and lower bounds on $N(d,d^\perp)$,
and also determine the true values of $N(d,d^\perp)$
for small $d$ and $d^\perp$.


\section{Upper Bound}


In this section,
we show upper bounds
on $N(d, d^{\perp})$.
The first bound is based on a Gilbert-Varshamov type
argument \cite[pp.~557--558]{macwilliams77}.

\begin{definition}
\begin{eqnarray*}
S_{n,m}&=&\{C \mid C \mbox{ is an }[n,m] \mbox{ linear code}\},\\
S_{n,m}(\myvec{v})&=&\{C\in S_{n,m} \mid C \ni \myvec{v}\},\\
S^\perp_{n,m}(\myvec{v})&=&\{C\in S_{n,m} \mid C^\perp \ni \myvec{v}\}
\end{eqnarray*}
\end{definition}

\begin{lemma}\label{lem3}
For a nonzero vector $\myvec{v}\in GF(2)^n$,
we have
\begin{eqnarray}
\frac{|S_{n,m}(\myvec{v})|}{|S_{n,m}|} &=&\frac{2^m-1}{2^{n}-1},\label{eq4}\\
\frac{|S_{n,m}^\perp(\myvec{v})|}{|S_{n,m}|} &=&\frac{2^{n-m}-1}{2^{n}-1}.\label{eq5}
\end{eqnarray}
\end{lemma}
Proof is given in Appendix \ref{app:lem}.

\begin{theorem}\label{upper1}
There exists an $[n,m,d]$ binary code with
the dual distance $d^\perp$ if
\[
\frac{2^{m}-1}{2^{n}-1}\sum_{i=1}^{d-1}{n \choose i} +
\frac{2^{n-m}-1}{2^{n}-1}\sum_{i=1}^{d^\perp-1}{n \choose i} < 1.
\]
$N(d,d^\perp)$ is upper bounded by the minimum $n$ 
satisfying the above inequality.
\end{theorem}

\begin{proof}
The required code exists iff
\[
S_{n,m} \neq \bigcup_{1\leq wt(\myvec{v}) \leq d-1} S_{n,m}(\myvec{v})
\cup \bigcup_{1\leq wt(\myvec{v}) \leq d^\perp-1} S^\perp_{n,m}(\myvec{v}).
\]
The cardinality of the right hand side is less than or equal to
\begin{eqnarray}
&&\sum_{1\leq wt(\myvec{v}) \leq d-1}|S_{n,m}(\myvec{v})| +
\sum_{1\leq wt(\myvec{v}) \leq d^\perp-1} |S^\perp_{n,m}(\myvec{v})|\label{eq10}\\
&\leq&\left(
\frac{2^{m}-1}{2^{n}-1}\sum_{i=1}^{d-1}{n \choose i} +
\frac{2^{n-m}-1}{2^{n}-1}\sum_{i=1}^{d^\perp-1}{n \choose i}
\right)|S_{n,m}| \nonumber
\end{eqnarray}
by Lemma \ref{lem3}. Thus, if the assumption of the theorem is satisfied,
the required code exists.
\end{proof}

We also introduce another upper bound.
\begin{proposition}
\begin{eqnarray}
N(d-1,d^\perp)&\leq&N(d,d^\perp)-1 \mbox{ (for $d\geq 2$)}, \label{up1}\\
N(d,d^\perp-1)&\leq&N(d,d^\perp)-1 \mbox{ (for $d^\perp\geq 2$)}. \label{up2}
\end{eqnarray}
\end{proposition}

\begin{proof}
Let $C$ be a linear code attaining $N(d,d^\perp)$,
and $C'$ be the punctured code of $C$.
Then $C'$ has the minimum distance at least $d-1$ and the dual distance
at least $d^\perp$, which proves Eq.~(\ref{up1}).
Equation~(\ref{up2}) is proved by considering the
punctured code of $C^\perp$.
\end{proof}

\section{Lower bounds}
In this section,
we give four lower bounds on $N(d, d^\perp)$.
The first two are immediate applications of
the Griesmer bound and a well-known fact of MDS codes.
The third is based on an improvement to
the Hamming bound.
The fourth is an improvement to
Brouwer's bound \cite{brouwer93} based on
the solvability
of a system of linear inequalities \cite{delsarte72}.

\subsection{Bounds based on the Griesmer bound and the result in MDS codes}
\begin{proposition}[Griesmer]
\cite[Section 17.\S 6]{macwilliams77}
If there exists an $[n,m,d]$ linear code,
then
\[
n \geq d+ \sum_{i=1}^{m-1} \left\lceil
\frac{d}{2^i}\right \rceil.
\]
\end{proposition}

\begin{theorem}\label{th:a}
\begin{eqnarray}
N(d,d^\perp) \geq
\lefteqn{\min [ n : 
2n \geq d+d^\perp +}\nonumber\\ &&
\hspace*{-1cm}\min_{m=1,\ldots,n-1} \left\{\sum_{i=1}^{m-1} \left\lceil
\frac{d}{2^i}\right \rceil +
\sum_{i=1}^{n-m-1} \left\lceil
\frac{d^\perp}{2^i}\right \rceil \right\}]\label{eq:gri}
\end{eqnarray}
\end{theorem}
\begin{proof}
If there exists an $[n,m,d]$ code with dual distance $d^\perp$,
then by the Griesmer bound we have
\begin{equation}
2n \geq d+d^\perp +
\sum_{i=1}^{m-1} \left\lceil
\frac{d}{2^i}\right \rceil +
\sum_{i=1}^{n-m-1} \left\lceil
\frac{d^\perp}{2^i}\right \rceil. \label{eq:d}
\end{equation}
Since $N(d,d^\perp)$ is the minimum $n$ such that
there exists a linear code of length $n$, minimum
distance $d$ and dual distance $d^\perp$,
$2N(d,d^\perp)$ is lower bounded by the minimum of
the right hand side of Eq.~(\ref{eq:d}) over possible
$n$ and $m$. \end{proof}

\begin{remark}
It is well-known that the simplex codes
attain the Griesmer bound.
However, they do not attain Eq.~(\ref{eq:gri}).
\end{remark}


The Singleton bound is a corollary to the Griesmer bound
and has a simpler expression. It
states that
if there exists an $[n,m,d]$ code then $m\leq n-d+1$.
When the code is binary and $d\geq 3$,
it can be tightened to $m\leq n-d$ \cite{pless98}.
The first part of the following result can be seen as
a corollary to Theorem~\ref{th:a}.

\begin{theorem}\label{coro:g}
\begin{equation}
N(d,d^\perp) \geq d+d^\perp -2. \label{eq:single1}
\end{equation}
When $d \geq 3$ and $d^\perp \geq 3$, we have\footnote{This improvement
was pointed out by an anonymous reviewer.}
\begin{equation}
N(d,d^\perp) \geq d+d^\perp.\label{eq:single2}
\end{equation}
\end{theorem}

\begin{proof}
Adding $m \leq N(d,d^\perp)-d+1$ and $N(d,d^\perp)-m \leq N(d,d^\perp)-d^\perp+1$
shows Eq.~(\ref{eq:single1}).
A similar argument shows Eq.~(\ref{eq:single2}). \end{proof}

\subsection{Bound based on an improved Hamming bound}
In this subsection,
we will introduce an improvement to the Hamming bound,
and derive a lower bound on $N(d,d^\perp)$ as
a corollary.

\begin{definition}
For positive integers $d$ and $n$,
we define the function $\ell(n,d)$ by
\[
\ell(n,d) =
\left\{
\begin{array}{ll}
\displaystyle \sum_{i=0}^{(d-1)/2} {n\choose i}&\mbox{for odd $d$},\\
\displaystyle \sum_{i=0}^{d/2-1}{n\choose i}
+ {n-1 \choose d/2-1}&\mbox{for even $d$}.
\end{array}\right.
\]
\end{definition}

Discrete random variables $X_1$, \ldots, $X_n$ are said to be
$d$-wise independent if
\[
\mathrm{Pr}[X_{i_1} = x_{i_1}, \ldots,
X_{i_d} = x_{i_d}] = \prod_{j=1}^d \mathrm{Pr}[X_{i_j} = x_{i_j}]
\]
for all $d$-tuples of indices $(i_1$, \ldots, $i_d)$
and all realizations $(x_{i_1}$, \ldots, $x_{i_d})$ of 
random variables.

\begin{lemma}\label{lem:b}\cite[Proposition 6.4]{alon86}
Let 
$X_1$, \ldots, $X_n$ be $(d-1)$-wise independent nonconstant
random variables
that map
the sample space $\Omega$ to $\{0,1\}$.
Then we have $|\Omega| \geq \ell(n,d)$.
\end{lemma}

\begin{theorem}\label{th:b}
For an $[n,m,d]$ linear code $C$,
we have $2^{n-m} \geq \ell(n,d)$.
\end{theorem}

\begin{proof}
Let $H$ be a parity check matrix for $C$,
and $h_i$ be its $i$-th column.
Consider the sample space $\Omega = GF(2)^{n-m}$
and the random variable $X_i$ that maps $v \in \Omega$
to the inner product of $v$ and $h_i$.
Since any $(d-1)$ columns in $H$ are linearly
independent, the random variables
$X_1$, \ldots, $X_n$ are $(d-1)$-wise independent
with the uniform probability distribution on $\Omega$.
By Lemma \ref{lem:b},
$2^{n-m} = |\Omega| \geq \ell(n,d)$.
\end{proof}

Observe that Theorem \ref{th:b} is an improvement
to the Hamming bound when $d$ is even.

\begin{corollary}\label{cor:g}
\[
N(d,d^\perp)\geq \min\{ n \mid n \geq \log_2 \ell(n,d) + \log_2 \ell(n,d^\perp)\}.
\]
\end{corollary}

\begin{proof}
If there exists an $[n,m,d]$ linear code with
dual distance $d^\perp$, then by Theorem \ref{th:b}
\begin{eqnarray}
&&2^{n-m}\cdot 2^m \geq \ell(n,d) \cdot \ell(n,d^\perp)\nonumber\\
&\Longleftrightarrow&
n \geq \log_2 \ell(n,d) + \log_2 \ell(n,d^\perp). \label{eq:e}
\end{eqnarray}
Since $N(d,d^\perp)$ is the minimum $n$ such that
there exists a linear code of length $n$, minimum
distance $d$ and dual distance $d^\perp$,
$N(d,d^\perp)$ is lower bounded by the minimum of
the right hand side of Eq.~(\ref{eq:e}) over possible
$n$. \end{proof}

\subsection{Bounds based on linear inequalities}\label{sec43}
For a linear code $C$, define
\begin{eqnarray*}
A_w &=& |\{ c \in C : wt(c) = w\}|,\\
A'_w &=& |\{ c \in C^\perp : wt(c) = w\}|.
\end{eqnarray*}
We have \cite[Section 5.\S 2]{macwilliams77}
\[
A'_w = \frac{1}{|C|}\sum_{i=0}^n A_i P_w(i)
= \frac{1}{|C|}{n \choose w} + \frac{1}{|C|}\sum_{i=1}^n A_i P_w(i),
\]
where
$P_w(i)$ is the Krawtchouk polynomial defined
by
\[
P_w(i) = \sum_{j=0}^w (-1)^j {i \choose j}{n-i \choose w-j}.
\]
For $w=1$, \ldots, $n$, we must have $A'_w \geq 0$.
When the code $C$ has minimum distance $d$,
we have $A_1 = A_2 = \cdots = A_{d-1} = 0$.
We also have $A'_1 = \cdots = A'_{d^\perp-1} = 0$
if $C$ has dual distance $d^\perp$.
Therefore,
if there exists a linear code of length $n$,
minimum distance $d$ and dual distance $d^\perp$,
then there exists a solution $A_d$, \ldots, $A_n$ to the following
system of linear inequalities:
\begin{equation}
\left\{\begin{array}{rcl}
A_i &\geq &0 \mbox{ for } i=d, \ldots, n,\\
\sum_{i=d}^n A_i P_w(i) &=& -{n\choose w}
\mbox{ for } w=1,\ldots,d^\perp-1,\\
\sum_{i=d}^n A_i P_w(i) &\geq& -{n\choose w}
\mbox{ for } w=d^\perp,\ldots,n.
\end{array}
\right. \label{eq:linear}
\end{equation}

\begin{theorem} \cite{brouwer93} \label{th:c}
$N(d,d^\perp)$ is greater than or equal to
the minimum $n$ such that there exists a solution
to the above system of linear inequalities.
\end{theorem}


We will add other constraints to Eq.~(\ref{eq:linear}).
Since we consider linear codes,
there must exist an \emph{integer} solution
$(A_d$, \ldots, $A_n)$ with $A_d + \cdots + A_n = 2^m - 1$
for some nonnegative integer $m$.

A binary linear code is said to be \emph{even}
if all codewords have even weight.
We call a code \emph{odd} if it is not even.
When the code $C$ is odd,
then there is the same number of even weighted codewords
and odd weighted ones. Moreover, the dual code $C^\perp$
does not contain the codeword with all $1$,
otherwise $C$ is even.
Therefore, if the code $C$ is odd, then we have
\begin{equation}
\left\{
\begin{array}{rcl}
\sum_{i: \textrm{even}} A_i &=& \sum_{i: \textrm{odd}} A_i,\\
A'_n &=& 0.
\end{array}\right.\label{eq:odd}
\end{equation}

When the code $C$ is even,
then the dual code $C^\perp$ contains 
the codeword with all $1$, and
we have $A'_i = A'_{n-i}$,
because there is one-to-one correspondence
between codewords with weight $i$ and weight $n-i$
by adding the all $1$ codeword.

Furthermore, we have the following inequality \cite{brouwer93}
when
$C$ is even:
\[
4 \sum_{4|i} A_i \geq \sum_{i=0}^n A_i,
\]
where $4|i$ denotes that $4$ divides $i$.
Summing up, the evenness of $C$ implies
\begin{equation}
\left\{
\begin{array}{rcl}
A_i &=& 0, \mbox{ for } i=1,3,5,\ldots,\\
4 \sum_{4|i} A_i &\geq& \sum_{i=0}^n A_i,\\
A'_n&=& 1,\\
A'_i &=& A'_{n-i}.
\end{array}\right.
\label{eq:even}
\end{equation}

By exchanging the role of $C$ and $C^\perp$,
we see that the oddness of $C^\perp$ implies
\begin{equation}
\left\{
\begin{array}{rcl}
\sum_{i: \textrm{even}} A'_i &=& \sum_{i: \textrm{odd}} A'_i,\\
A_n &=& 0.
\end{array}\right.\label{eq:dualodd}
\end{equation}
and that the evenness of $C^\perp$ implies
\begin{equation}
\left\{
\begin{array}{rcl}
A'_i &=& 0, \mbox{ for } i=1,3,5,\ldots,\\
4 \sum_{4|i} A'_i &\geq& \sum_{i=0}^n A'_i,\\
A_n&=& 1,\\
A_i &=& A_{n-i}.
\end{array}\right.
\label{eq:dualeven}
\end{equation}

When we estimate $N(d,d^\perp)$ and $d$ is even,
the code can be either odd or even,
and we search a solution for either Eq.~(\ref{eq:odd}) or
(\ref{eq:even}). When $d$ is odd, the code is odd and
we search a solution for Eq.~(\ref{eq:odd}) only.
The same rule applies to $d^\perp$.

\begin{remark}
We remark on the computational complexity on the bound presented
in this subsection.
When we require $A_d$, \ldots, $A_n$ to be integers,
we have to solve an integer programming problem for which there is
no known polynomial time algorithm in the number of variables
\cite[Section~11.8]{bertsimas97}.
When we allow $A_d$, \ldots, $A_n$ to be any real numbers,
we solve a linear programming problem that can be solved
in roughly $O((n-d)^5)$ arithmetic operations \cite[Section~9.3]{bertsimas97}.
In both case, it quickly becomes difficult to compute the lower bound
for large $n$.
\end{remark}

\subsection{Numerical Examples}
In this subsection,
we give numerical examples of the derived bounds
in Table \ref{tab1}.
An entry $x$ in Table \ref{tab1} means
that $N(d,d^\perp) \geq x$ for the lower bounds,
and $N(d,d^\perp) \leq x$ for the upper bound.
True values of $N(d,d^\perp)$ are also listed,
which are obtained by  exhaustive search.
Generator matrices of codes attaining $N(d,d^\perp)$
are listed in Appendix \ref{app:mat}.
We could not determine the true values of $N(d,d^\perp)$
by exhaustive search with $(d,d^\perp)$ not listed in Table \ref{tab1}.
We remark that $N(2,\delta) = N(\delta,2) = \delta$
because the trivial $[\delta, 1, \delta]$ code has dual distance $2$.

{From} Table \ref{tab1},
we can make the following observations.
Lower bounds are increasing in order of
Corollary~\ref{cor:g}, Theorem~\ref{th:c}, and the improvement
of Theorem~\ref{th:c} in Sec.~\ref{sec43}.
Theorems~\ref{th:a} and \ref{coro:g} give smaller lower bounds.
The upper bound in Theorem~\ref{upper1} is very loose for small values of
$d$ and $d^\perp$.
This looseness seems to come from the fact that
many elements are counted several times in Eq.~(\ref{eq10}).

Additional constraints in Sec.~\ref{sec43} give the true values
of $N(d,d^\perp)$ as a lower bound except for $(d,d^\perp) = (5,5)$.
They also improve Theorem~\ref{th:c} in the parameters
$(d,d^\perp) = (5,3)$, $(5,4)$, $(6,3)$, $(6,4)$, $(6,5)$, $(6,6)$.
These improvements significantly reduced the required time
for exhaustive search.

\begin{table*}
\caption{True values and estimates of $N(d,d^\perp)$ by the derived bounds}\label{tab1}
\begin{center}
\begin{tabular}{|c|c||c||c|c|c|c|c||c|}\hline
$d$&$d^\perp$&\multicolumn{1}{p{6ex}||}{true value}&
\multicolumn{5}{c||}{lower bounds}&\multicolumn{1}{p{8ex}|}{upper bound}\\\cline{4-9}
&&&Thm.~\ref{coro:g}&Thm.~\ref{th:a}&Cor.~\ref{cor:g}&\multicolumn{1}{p{7ex}|}{Thm.~\ref{th:c} (conventional)}&Sect.~\ref{sec43}&Thm.~\ref{upper1}
\\\hline\hline
3&3&6&6&5&6&6&6&17\\ 
4&3&7&7&6&7&7&7&21\\ 
4&4&8&8&7&8&8&8&25\\ 
5&3&11&8&7&9&10&11&24\\ 
5&4&13&9&8&11&11&13&29\\ 
5&5&16&10&11&14&14&14&34\\ 
6&3&12&9&8&10&11&12&28\\ 
6&4&14&10&9&12&12&14&33\\ 
6&5&17&11&12&15&15&17&38\\ 
6&6&18&10&13&16&16&18&42\\ 
7&3&14&10&9&12&14&14&31\\ 
7&4&15&11&10&14&15&15&37\\ 
8&3&15&11&10&14&15&15&35\\ 
8&4&16&12&11&15&16&16&40\\\hline 
\end{tabular}
\end{center}
\end{table*}

\section{Conclusion}
In this paper,
we considered the minimum length of linear codes
with specified minimum Hamming distances and dual distances,
from which cryptographic Boolean functions are constructed.
We obtained an upper bound by a Gilbert-Varshamov type argument,
and lower bounds by applying the Griesmer, the Hamming, and
the linear programming bound.
The true values for the minimum length
are also determined by exhaustive search for certain range
of parameters.
These lower bounds and true values are useful for estimating
the necessary input length of cryptographic Boolean
functions for given cryptographic strength.
This paper also demonstrated that the upper bound proposed
herein is too loose, and it remains an open problem to
derive a tight upper bound.

\section*{Acknowledgment}
We would like to thank the reviewers' critical comments
which improve this paper a lot. In particular,
Theorem~\ref{coro:g} is improved by the reviewer's comment.

\appendices
\section{Proof of Lemma \ref{lem3}}\label{app:lem}
\begin{lemma}
For nonzero vectors 
$\myvec{u}$,
$\myvec{v}\in GF(2)^n$,
we have
\begin{eqnarray}
|S_{n,m}(\myvec{u})| &=& |S_{n,m}(\myvec{v})|,\label{eq1}\\
|S^\perp_{n,m}(\myvec{u})| &=& |S_{n,n-m}(\myvec{u})|,\label{eq2}\\
|S^\perp_{n,m}(\myvec{u})| &=& |S_{n,m}^\perp(\myvec{v})|.\label{eq3}
\end{eqnarray}
\end{lemma}

\begin{proof}
We define the group $GL_n$
as the set of bijective linear maps $f$ on $GF(2)^n$.
In the following equation,
$S_{n,m} \ni C_1$ is a fixed linear code,
and $g$ is a fixed bijective linear map on $GF(2)^n$
such that $g(\myvec{v})=\myvec{u}$.
\begin{eqnarray*}
&&|S_{n,m}(\myvec{u})|\\
&=&|\{C\in S_{n,m} \mid C \ni \myvec{u}\}|\\
&=&|\{f(C_1) \mid f(C_1) \ni \myvec{u}, f \in GL_n\}|\\
&=&|\{f(C_1) \mid f(C_1) \ni g(\myvec{v}), f \in GL_n\}|\\
&=&|\{g^{-1}\circ f(C_1) \mid g^{-1}\circ f(C_1) \ni \myvec{v}, f \in GL_n\}|\\
&=&|\{ f(C_1) \mid  f(C_1) \ni \myvec{v}, f \in GL_n\}|\\
&=&|S_{n,m}(\myvec{v})|.
\end{eqnarray*}
Equation~(\ref{eq1}) is proved.

By taking the dual code, we see that there is a one-to-one correspondence between
$S_{n,m}$ and $S_{n,n-m}$, and we have
\begin{eqnarray*}
&&|S^\perp_{n,m}(\myvec{u})|\\
&=&|\{C\in S_{n,m} \mid C^\perp \ni \myvec{u}\}|\\
&=&|\{C\in S_{n,n-m} \mid C \ni \myvec{u}\}|\\
&=&|S_{n,n-m}(\myvec{u})|,
\end{eqnarray*}
which proves Eq.~(\ref{eq2}).
Equation~(\ref{eq3}) is deduced from Eqs.~(\ref{eq1}) and (\ref{eq2}).
\end{proof}

\noindent\emph{Proof of Lemma \ref{lem3}}.
Let $B$ be the set of a pair of a nonzero  vector $\myvec{u}$
and $C\in S_{n,m}$ such that $\myvec{u}\in C$.
For each $C\in S_{n,m}$, there are $2^m-1$ nonzero vectors
$\myvec{u}$ such that $\myvec{u}\in C$, and we have $|B| =
(2^m-1)|S_{n,m}|$.

For each nonzero vector $\myvec{u}$ there are $|S_{n,m}(\myvec{u})|$
linear codes $C$ such that $\myvec{u}\in C$, and we have
\[
|B| = \sum_{\mathbf{0}\neq \myvec{u}\in GF(2)^n}
|S_{n,m}(\myvec{u})| = (2^n-1)|S_{n,m}(\myvec{v})|
\]
by Eq.~(\ref{eq1}). Thus Eq.~(\ref{eq4}) is proved.
Equation (\ref{eq5}) follows from Eqs.~(\ref{eq2}) and (\ref{eq4}).
\qed

\section{Linear codes attaining $N(d,d^\perp)$}\label{app:mat}
In this Appendix,
we give the name or generator matrices of linear codes attaining
$N(d,d^\perp)$. Matrices are generator matrices of linear codes
attaining $N(d,d^\perp)$ unless otherwise specified.

\medskip
\noindent
$N(3,3) = 6$: Attained by the $[6,3,3]$ shortened Hamming code.\\
$N(4,3)=7$: Attained by the $[7,4,3]$ Hamming code.\\
$N(4,4)=8$: Attained by the $[8,4,4]$ extended Hamming code.\\
$N(5,3)=11$:
\[
\left(
\begin{array}{ccccccccccc}
1&0&0&0&0&0&0&1&1&1&1\\
0&1&0&0&0&1&1&0&0&1&1\\
0&0&1&0&1&0&1&0&1&0&1\\
0&0&0&1&1&1&0&1&0&1&0
\end{array}\right)
\]
$N(5,4)=13$:
\[
\left(
\begin{array}{ccccccccccccc}
1&0&0&0&0&0&0&0&0&1&1&1&1\\
0&1&0&0&0&0&1&1&1&0&0&0&1\\
0&0&1&0&0&1&0&1&1&0&1&1&0\\
0&0&0&1&0&1&1&0&1&1&0&1&0\\
0&0&0&0&1&1&1&1&0&1&1&0&1
\end{array}\right)
\]
$N(5,5)=16$:
\[
\left(
\begin{array}{cccccccccccccccc}
1&0&0&0&0&0&0&0&0&0&0&0&1&1&1&1\\
0&1&0&0&0&0&0&0&0&0&1&1&0&0&1&1\\
0&0&1&0&0&0&0&0&0&1&0&1&0&1&0&1\\
0&0&0&1&0&0&0&0&0&1&1&0&1&0&1&0\\
0&0&0&0&1&0&0&0&1&0&0&1&0&1&1&0\\
0&0&0&0&0&1&0&0&1&0&1&0&1&0&1&1\\
0&0&0&0&0&0&1&0&1&1&0&1&1&0&1&1\\
0&0&0&0&0&0&0&1&1&1&1&0&1&1&0&1
\end{array}\right)
\]
$N(6,3)=12$:
\[
\left(\begin{array}{cccccccccccc}
1&0&0&0&0&0&0&1&1&1&1&1\\
0&1&0&0&0&1&1&0&0&1&1&1\\
0&0&1&0&1&0&1&0&1&0&1&1\\
0&0&0&1&1&1&0&1&0&1&0&1
\end{array}\right)
\]
$N(6,4)=14$: The generator matrix of its dual code is
\[
\left(\begin{array}{cccccccccccccc}
1&0&0&0&0&0&0&0&0&0&0&1&1&1\\
0&1&0&0&0&0&0&0&0&0&1&0&1&1\\
0&0&1&0&0&0&0&0&0&0&1&1&0&1\\
0&0&0&1&0&0&0&0&0&0&1&1&1&0\\
0&0&0&0&1&0&0&0&0&1&0&0&1&1\\
0&0&0&0&0&1&0&0&0&1&0&1&0&1\\
0&0&0&0&0&0&1&0&0&1&0&1&1&0\\
0&0&0&0&0&0&0&1&0&1&1&0&0&1\\
0&0&0&0&0&0&0&0&1&1&1&0&1&0
\end{array}\right)
\]
$N(6,5)=17$: The generator matrix of its dual code is
\[
\left(\begin{array}{ccccccccccccccccc}
1&0&0&0&0&0&0&0&0&0&0&0&1&1&1&1&1\\
0&1&0&0&0&0&0&0&0&0&1&1&0&0&1&1&1\\
0&0&1&0&0&0&0&0&0&1&0&1&0&1&0&1&1\\
0&0&0&1&0&0&0&0&0&1&1&0&1&0&1&0&1\\
0&0&0&0&1&0&0&0&1&0&0&1&0&1&1&0&1\\
0&0&0&0&0&1&0&0&1&0&1&0&1&0&1&1&0\\
0&0&0&0&0&0&1&0&1&1&0&1&1&0&1&1&1\\
0&0&0&0&0&0&0&1&1&1&1&0&1&1&0&1&1
\end{array}\right)
\]
$N(6,6)= 18$:
\[
\left(
\begin{array}{cccccccccccccccccc}
1&0&0&0&0&0&0&0&0&0&0&0&0&1&1&1&1&1\\
0&1&0&0&0&0&0&0&0&0&0&1&1&0&0&1&1&1\\
0&0&1&0&0&0&0&0&0&0&1&0&1&0&1&0&1&1\\
0&0&0&1&0&0&0&0&0&0&1&1&0&1&0&1&0&1\\
0&0&0&0&1&0&0&0&0&1&0&0&1&0&1&1&0&1\\
0&0&0&0&0&1&0&0&0&1&0&1&0&1&0&1&1&0\\
0&0&0&0&0&0&1&0&0&1&1&0&1&1&0&1&1&1\\
0&0&0&0&0&0&0&1&0&1&1&1&0&1&1&0&1&1\\
0&0&0&0&0&0&0&0&1&1&1&1&1&0&1&1&1&0
\end{array}\right)
\]
$N(7,3)=14$: Attained by the $[14,4,7]$ punctured simplex code.\\
$N(7,4)=15$: Attained by the $[15,5,7]$ punctured first order Reed-Muller code.\\
$N(8,3)=15$: Attained by the $[15,4,8]$ simplex code.\\
$N(8,4)=16$: Attained by the $[16,5,8]$ first order Reed-Muller code.


%



\end{document}